\documentclass{aastex}
\usepackage{spr-astr-addons}
\def\l{$\lambda$}
\def\mbh{$M_{\rm BH}$\/}

\def\lledd{$L/L_{\rm Edd}$}

\def\rfe{$R_{\rm FeII}$\/}

\def\feiiq{\rm Fe{\sc ii}$\lambda$4570\/}
\def\msol{M$_\odot$\/}

\def\ltsima{$\; \buildrel < \over \sim \;$}
\def\ltsim{\lower.5ex\hbox{\ltsima}}  
\def\simlt{\lower.5ex\hbox{\ltsima}}  
\def\gtsima{$\; \buildrel > \over \sim \;$}

\def\gtsim{\lower.5ex\hbox{\gtsima}} 
\def\simgt{\lower.5ex\hbox{\gtsima}}

\def\civ{{\sc{Civ}}$\lambda$1549\/}

\def\cmq{cm$^{-2}$\/}
\def\cm3{cm$^{-3}$\/}
\def\hb{{\sc{H}}$\beta$\/}

\def\hbnc{{\sc{H}}$\beta_{\rm NC}$\/}

\def\oiiiopt{{\sc{[Oiii]}}\-$\lambda\lambda$\-4959,\-5007\/}
\def\o4363{{\sc{[Oiii]}}$\lambda$4363\/}

\def\feii{{Fe\sc{ii}}\/}

\def\fe{{\sc{Fe}}\/}
\def\dvr{{$\Delta$ v$_r$}}

\def\fe76087{{\sc [Fe vii]}$\lambda$6087\/}
\def\oiii{{\sc [Oiii]}$\lambda$5007}

\def\kms{km~s$^{-1}$}

\def\ergss{ergs s$^{-1}$\/}

\RequirePackage{color}

\begin{document}

\title{BLUE OUTLIERS AMONG { INTERMEDIATE} REDSHIFT QUASARS}
\shorttitle{Short article title}
\shortauthors{Autors et al.}

\author{P. Marziani}
\affil{INAF, Osservatorio Astronomico di Padova, Padova, Italia}
\and 
\author{J. W. Sulentic}
\affil{Instituto de Astrof{\'\i}sica de Andaluc{\'\i}a (CSIC),    Granada, Spain}
\and
\author{G. M. Stirpe}
\affil{INAF, Osservatorio Astronomico di Bologna, Bologna, Italia}
\and
\author{D. Dultzin}
\affil{Instituto de Astronom{\'\i}a, UNAM, 
Mexico, D.F.,   Mexico}
\and
\author{A. Del Olmo}
\affil{Instituto de Astrof{\'\i}sica de Andaluc{\'\i}a (CSIC),    Granada, Spain}
\and
\author{M. A. Mart\'{\i}nez-Carballo}
\affil{Instituto de Astrof{\'\i}sica de Andaluc{\'\i}a (CSIC),    Granada, Spain}



\defcitealias{zamanovetal02}{Z02}
\begin{abstract}
\oiii\ ``blue outliers" -- that are suggestive of outflows in the  narrow
line region of quasars --  appear to be much more common 
at intermediate $z$\ (high luminosity)  than at low $z$.    About $40$\%\
of quasars in a Hamburg ESO intermediate $z$ sample of 52 sources qualify
as ``blue outliers"  (i.e., quasars with [OIII] $\lambda\lambda$4959,5007 
lines showing large systematic blueshifts with respect to rest frame).
We discuss major findings  on what has become an intriguing field in
active galactic nuclei research and  stress the relevance of ``blue
outliers" to  feedback  and host galaxy evolution.
\end{abstract}


\section{Introduction}
\label{intro} 
Low redshift quasars usually show both broad and narrow emission lines in their optical spectra. The type-1/type-2 distinction arose because some quasars showed only narrow emission lines in (unpolarized) light. Studies of type-1 quasars over the last 20-30 years have shown
that the most studied \oiii\ narrow line involves: 1) a strong
unshifted narrow and 2) a usually weaker blueshifted and slightly broader
component. Blue outliers   were originally defined as quasars having  their [OIII] $\lambda\lambda$4959,5007  emission lines  blueshifted by more than 250 km s$^{-1}$\ at peak \citep{zamanovetal02,marzianietal03b}. They were thought to be rare sources, perhaps too rare to affect the measurement of AGN systematic redshifts: optical narrow lines have been   used for decades  to estimate rest-frame systemic $z$\ of quasars with little attention for the ionization stage of the emitting ionic species. However, a systematic radial velocity difference between narrow  low-ionization  and high ionization lines (like [OIII] $\lambda\lambda$4959,5007) is now established  \citep{zamanovetal02,eracleoushalpern04,huetal08}. The distributions { of peak line shifts} of \citet[][hereafter \citetalias{zamanovetal02}]{zamanovetal02} and \citet{huetal08} are remarkably similar, with the mode of the \citet{huetal08} data displaced by $\approx$ --30 km s$^{-1}$\   with respect to  [OII]$\lambda$3727.  Both distributions also show a skew toward \oiii\ blueshifts. A striking property of the \citetalias{zamanovetal02} distribution is the presence of  unexpectedly large blueshifts: in the histogram they appear  detached from the main distribution.  It is not difficult to show that these sources are outliers following standard statistical tests, for example the one based on modified $Z$ score { applied to the peak velocity shift $v_\mathrm{i}$\ of source $i$} \citep{iglewiczhoaglin93}: 
\begin{equation}
Z_\mathrm{i} = 0.6745 \frac{v_{i} - <v>}{ <|v_{i} - <v>|>}
\end{equation}
where the brackets indicate the median operator and the term at the denominator is the median absolute deviation. Blue outliers (BOs) have since   been found in several samples \citep{bianetal05,aokietal05, komossaetal08,zhangetal11,zhangetal13}. However, until 2008  (\citealt{komossaetal08} included)     only 27 sources   qualified as BOs  according to the definition of \citetalias{zamanovetal02}. In the samples of \citet{bianetal05} and of \citet{marzianietal03a} they were $\approx$ 3 \% and  5\%\ of all sources, respectively: rare sources, but not extremely rare, with the precise amount of their prevalence being obviously sensitive to sample selection criteria. 

The first issue in measurements of  the [OIII] $\lambda\lambda$\-4959,\-5007 shift is to set a reliable rest frame reference.   Ideally, the rest frame should be set by the systemic redshift of the host galaxy, from HI or CO lines, or stellar absorption lines  \citepalias[][\citealt{baewoo14}]{zamanovetal02}. More expedient in the analysis of large ensembles of quasar spectra is to estimate the quasar redshift from low-ionization emission lines (typically H$\beta$ or H$\alpha$\ narrow component, the partially resolved [OII]$\lambda$3727 doublet, or [NII]$\lambda$6548,6583). There are several caveats also with this approach but these lines are customarily sharply peaked and, with good S/N ($\gtsim 20$), make possible the definition of a conventional  quasar rest frame with a precision $\approx \pm 50$ km s$^{-1}$\   \citep{marzianietal13a}. 

Progressing to high redshift is not easy. The equivalent width of  [OIII] $\lambda\lambda$4959,5007 strongly decreases as a  function of redshift and/or luminosity \citep[e.g.][]{brothertonetal96,netzeretal04,baskinlaor05c,zhangetal13}  The \oiii\ line is often of low equivalent width (few \AA), and relatively broad ($\approx$ 1000 km s$^{-1}$, \citealt{netzeretal04}).  Therefore it does  not come as a surprise that little data exists for [OIII] $\lambda\lambda$4959,5007 shifts at $z > 1$: the lines are shifted into the  near IR (NIR), and spectroscopic observation in the JHK bands were limited to the brightest quasars until very recent times. 

In this paper we report on  the observation of the [OIII] $\lambda\lambda$4959,5007 line properties in a sample of 52 intermediate  redshift ($z \gtsim 1$) quasars. { The main aim is to operate a comparison between the [OIII] $\lambda\lambda$4959,5007 emission line properties at low-$z$\ and the luminous quasars at $z \gtsim 1$. The intermediate $z$\  sample involves sources that are among the most luminous quasars, observed at a cosmic epoch ($z \sim 1 - 2$) when the quasar population was reaching its maximum ``splendour'' in terms of space density and luminosity  \citep[e.g.,][]{boyleetal00,richardsetal06,walletal05}.   BOs might be associated with nuclear outflows but they are relatively infrequent at low-$z$\ (\S \ref{e1lowz}). Until now, direct evidence of mechanical feedback from the active nuclei on their host galaxies has been rather elusive at low-$z$. However, feedbacks effects are expected to have been operating at higher redshift for massive galaxies, at least in order to account for the \mbh\ -- $M_{\star}$\ correlation observed among local Universe galaxies \citep[][]{merritt01,gebhardtetal00}. }   It might well be possible that  outflows associated with \oiiiopt\ blueshifts  were more frequent in the past. Their energetics  may provide evidence in favour of  feedback effects on the quasars' host galaxies. In this paper we first   summarize the basic aspects that have emerged by the study of    \oiiiopt\ and specifically of the BOs in the Eigenvector 1 context at low-$z$\ (\S \ref{e1lowz}). We then summarily describe observations and data analysis of the quasar spectra  at intermediate redshift (\S \ref{obs}). Main results concerning the \oiiiopt\ line profile are reported  in \S  \ref{res} and discussed in \S \ref{disc} { where we show that intermediate-$z$\ BOs are frequent  at high luminosity and may be associated with   high-kinetic power outflows.} 

\section{Blue outliers in the context of Eigenvector 1 at low-$z$}
\label{e1lowz}

It is known since the early 1980s that the   [OIII]$\lambda\lambda$\-4959,\-5007 profiles can be strongly asymmetric with a predominance of blue-ward asymmetries \citep[e.g.,][]{whittle85,vrtilekcarleton85}, but a more focused view emerged with the analysis of [OIII]$\lambda\lambda$4959,5007 in the Eigenvector 1 context \citep{borosongreen92,sulenticetal00a,zamanovetal02,marzianietal03b}. {E1 was first defined by \citet{borosongreen92} through a principal component analysis of { 87} quasars. The original E1 was dominated by an anti correlation between \oiii\ peak intensity or and FeII prominence, and was later extended into a 4D E1 by \citet{sulenticetal00b} involving parameters associated with broad line and continuum emission, namely FWHM of broad \hb, \feii\ prominence parameter, } \rfe = \feiiq/\hb; { \civ\ centroid shift at one-half fractional intensity, and soft X-ray photon index. Of special importance is the optical plane of E1 defined by FWHM(\hb) vs. \rfe. In this plane data points from large samples of  quasars can be plotted provided that a high S/N spectrum is available { for each source} in the \hb\ spectral range \citep{sulenticmarziani15}. \citet{sulenticetal00a} identified two populations, A (FWHM $\le 4000$ \kms) and B (FWHM $> 4000$ \kms). Composite spectra can be build for spectral bins from A1 to A4, in order of increasing \rfe, from 0 in steps of $\Delta$\rfe=0.5 (for example A2 has $0.5 \le $ \rfe $< 1.0$); A1 to B1 to B1$^{++}$\ in order  of increasing \hb\ FWHM, with $\Delta$FWHM = 4000 \kms\ \citep{sulenticetal02}. }  In the optical plane of E1, large equivalent width W([OIII]$\lambda\lambda$4959,5007) sources can have a blue-ward asymmetric wing; low W([OIII]$\lambda\lambda$4959,5007) sources (located in the lower left part of the optical E1 diagram) tend to have emission dominated by  a semi-broad (1000 -- 2000 km s$^{-1}$), most-often blueshifted feature that may resemble the blueward wing observed in stronger [OIII]$\lambda\lambda$4959,5007 emitters \citep[e.g.,][]{marzianietal03b,marzianietal06,zhangetal13}. 

The BOs occupy the lower right part of the diagram and are therefore exclusively Pop. A/narrow line Seyfert 1 (NLSy1)  nuclei (\citetalias{zamanovetal02}). Use of the \oiiiopt\ line width as a proxy of the {bulge stellar component} velocity dispersion should be avoided for population A sources in the spectral types A2, A3, A4 \citep[e.g.,][and references therein]{marzianisulentic12}. The narrow line region (NLR) in sources whose \oiiiopt\ profile is blueshifted and semi-broad is unlikely to be dynamically related to the host-galaxy stellar bulge. \citet{marzianisulentic12} suggest a minimum equivalent width $\approx$ 20 \AA\ to ensure dominance by the narrower, unshifted (or less shifted) core component; otherwise, the \oiiiopt\ FWHM may lead to a significant overestimation of black hole masses \citep{botteetal05}.  Large blueshift and the profile of the [OIII]$\lambda\lambda$4959,5007 lines are explained as due to a radial outflow \citep[\citetalias{zamanovetal02}, ][]{komossaetal08}. \citetalias{zamanovetal02} were able to account for the \oiii\ line profile assuming a wide angle radial outflow with a velocity field  $v(r) = {\cal F}v_\mathrm{esc}(r) \propto r^{-\frac{3}{4}}$, i.e., with gas moving radially at a fixed fraction ${\cal F}>1$ of the local escape velocity at $r$, $v_\mathrm{esc}(r)$.   The low W(\oiiiopt) is also consistent with (but not a proof of) a compact NLR. {Evidence exists that low-$z$ BOs involve a nuclear outflow. We will show that this is likely the case also at intermediate $z$\ (\S \ref{res} and \ref{disc}).}  

\begin{table*}[htp]
\caption{Sample properties}
\label{tab:bo}       
\setlength{\tabcolsep}{4pt}
\begin{tabular}{lcccccccccc}
\hline\noalign{\smallskip}
Sample & $N_\mathrm{tot}$ & $N_\mathrm{A}$ &$z_\mathrm{sample}$ & $M_\mathrm{abs}$ &  $N_\mathrm{BO}$ & Shift & FWHM & $f_\mathrm{NLSy1}^\mathrm{a}$ & $f_\mathrm{A}^\mathrm{b}$  \\
&&&&&& [\kms] & [\kms] &\\
\noalign{\smallskip}\hline\noalign{\smallskip}
\citet{marzianietal03a} & 215 & 95 & 0.0 -- 0.7  &  --27 /-- 21  & 7 &  $\approx$ --2500 / --250 & 500 -- 2000 &  18 \% & 7\%\\
\citet{bianetal05} &          150 & 150 & 0.1 -- 0.6 & --25 / --21 & 7  &$\approx$ --1000 / --250   & 400 -- 1300 &   5\% & 5\% \\
HE (present work) &   52  & 27 & 1.0 -- 2.2  & --30 / --26.5 & 21 & $\approx$ --2500 / --250  & 500 -- 3000 & 0\%  & 48\%\ \\
\citet{netzeretal04} &  29     &   16$^\mathrm{c}$    & 2.0  -- 2.5$^\mathrm{c}$  & --28.5  / --27$^\mathrm{c}$                &       \ldots         & \ldots      &  600 -- 1800 & \ldots & \ldots  \\
\noalign{\smallskip}\hline
\end{tabular}
\, \,  $^\mathrm{a}$\ Satisfying the condition FWHM(\hb) $\le$ 2000 \kms. $^\mathrm{b}$\ Following the classification of \citet{marzianietal09}. \,  $^\mathrm{c}$\ Data retrieved from \citet{shemmeretal04}. Pop. A sources identified following \citet{marzianietal09}.
\end{table*}

\section{Observations and data analysis}
\label{obs}

ISAAC observations for 52  quasars from the Hamburg-ESO survey \citep[][hereafter HE sample]{wisotzkietal00} have been presented and discussed in \citet{sulenticetal04}, \citet{sulenticetal06}, and \citet{marzianietal09}. The data were obtained with low dispersion  grisms  and a narrow slit (0.6 arcsec) that ensured a spectroscopic resolution $\lambda/\Delta \lambda \approx$ 1000. Sources are in the redshift range  $0.9 < z <2.2$, with one source at $z \approx 3$\ (HE 0940--1050). The bolometric luminosity is in the range $47 < \log L < 48.5$ \ [\ergss]: the quasars in the HE sample are among the most luminous quasars known.

\begin{figure}[htp!]
\includegraphics[scale=0.35]{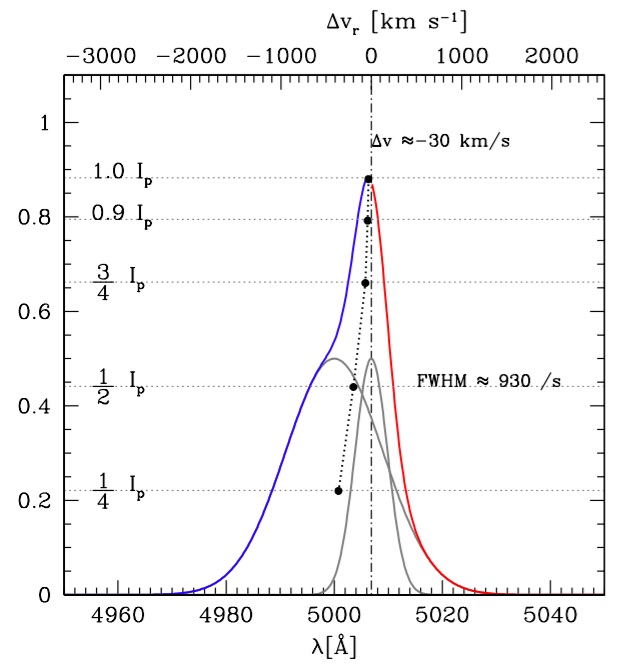}
\caption{Mock \oiii\ profile to illustrate the line decomposition into a core and a semibroad component. { Abscissa is rest-frame wavelength (bothom) and radial velocity from rest wavelength of \oiii\ (top). Ordinate is arbitrary intensity. 
The thick line shows the full profile (blue if \dvr $<0$, red if \dvr $<0$) due to the sum of the semibroad and core components (grey lines). }
Radial velocities are measured for the peak intensity $I_\mathrm{p}$\ ($\Delta v \approx$ 30 \kms),  and the centroids (black spots) at different fractional intensities.  }
\label{fig:mock}        
\end{figure}

A suitable low-$z$, low-$L$  control sample is offered by the 215 sources of \citet{marzianietal03a} since spectra were collected with a comparable resolution and S/N distribution, and were analyzed following a similar technique. { The  \citet{marzianietal03a} dataset is a rather heterogeneous collection of high S/N spectra that is loosely representative of the low-$z$\ type-1 sources and includes luminous Seyfert-1 in addition to quasars.  A larger sample, selected more rigorously from a well-defined flux limit \citep{zamfiretal10} shows an occupation in the optical plane of 4DE1 similar to \citet{marzianietal03a} but unfortunately has no \oiiiopt\ measures available. Reliable \oiiiopt\ measurements are also available  for the  sample of   \citet{bianetal05} that includes 150 NLSy1s  selected from the DR3 of the SDSS { where BOs should be more frequently found (\S \ref{distr}).} { Table \ref{tab:bo} summarizes the basic properties of the two low-$z$ and of the HE samples. Columns list, in this order, sample identification, total number of objects in sample, number of Pop. A objects, the sample redshift and absolute magnitude range, the number of BOs, an indicative shift and FWHM range, and the fractions of BOs that are NLSy1s and that belong to Pop. A. Range values are indicative of the bulk of redshift, magnitude and shift distributions in the various samples.  A helpful, high luminosity comparison sample is also provided by \citet{netzeretal04} who however yielded \oiiiopt\ equivalent width and FWHM measures but no  shifts.  The \citet{netzeretal04} is the only reasonably-sized sample with an uniform data-set in the same $L$ range that can be used for a comparison with the HE measures. The HE and \citet{netzeretal04} samples are representative of   high luminosity end of the quasar luminosity function at $z \approx$ 2. They were rare then, and completely absent at low $z$ ($z \simlt 0.7$, \citealt{boyleetal00,richardsetal06}).  

} 

The HE spectra   were modeled with a $\chi^{2}$\ minimization procedure using the routine {\sc specfit} implemented in the IRAF package \citep{kriss94}. The details of the analysis techniques have been described elsewhere \citep[][]{marzianietal09,marzianietal13a,negreteetal14}.  Here we recall that the main elements of the multi-component fits are continuum shape (a power-law), \feii\ emission template, a Lorentzian function to represent Pop. A H$\beta$ profiles, two Gaussians (one broad and redshifted) for   Pop. B H$\beta$. The  \oiiiopt\ lines were modeled assuming that each line is the sum of two Gaussians, one narrower and one broader, with the following constraints: same shifts, same FWHM, and intensity ratio fixed by atomic physics for \oiii\ and [O{\sc iii}]\l 4959 \citep{dimitrijevicetal07}. This approach is meant to reproduce a component presumably associated with the outer NLR (whose broadening is set by the bulge potential) and one component whose broadening is associated with non-virial motions. Minimum  $\chi_{\nu}^{2}$\ values are usually consistent with a satisfactory model of the    \oiii\ profile, although the relative intensity of the two Gaussians  (narrow and semi-broad) may not be well constrained. Also, in several cases, it was found that a single, shifted semibroad component accounted for the full profile. Line shifts and width at fractional peak intensity of $\frac{1}{4}$, $\frac{1}{2}$, $\frac{3}{4}$\ and 0.9 were also computed on the full profile to bypass the model dependent profile decomposition into two Gaussians. Fig. \ref{fig:mock} shows the decomposition for one typical case with a modest net blueshift at peak. Low W([OIII]$\lambda\lambda$4959,5007) and highly blueshifted profiles may show a much weaker core component or even only the single blue-shifted component. 

Line shifts were computed with respect to the H$\beta$ narrow component, or with respect to the peak of H$\beta$ is no narrow component was clearly identifiable. This procedure is safe for the detection of large  shifts since: 1) for Pop. A sources, the H$\beta$ broad profile is most often Lorentzian-like, and spiky. Shifts between narrow and broad component have been found to be less than 200 \kms\ \citep{sulenticetal12}; 2) broader Pop. B H$\beta$\ profiles show stronger H$\beta$ NC, and in general, stronger emission lines. Wavelength calibration residuals are not relevant, since they were  always  $\Delta\lambda /\lambda \ll 10^{-4}$.   

\section{Results}
\label{res}

\begin{figure}
 \includegraphics[scale=0.09]{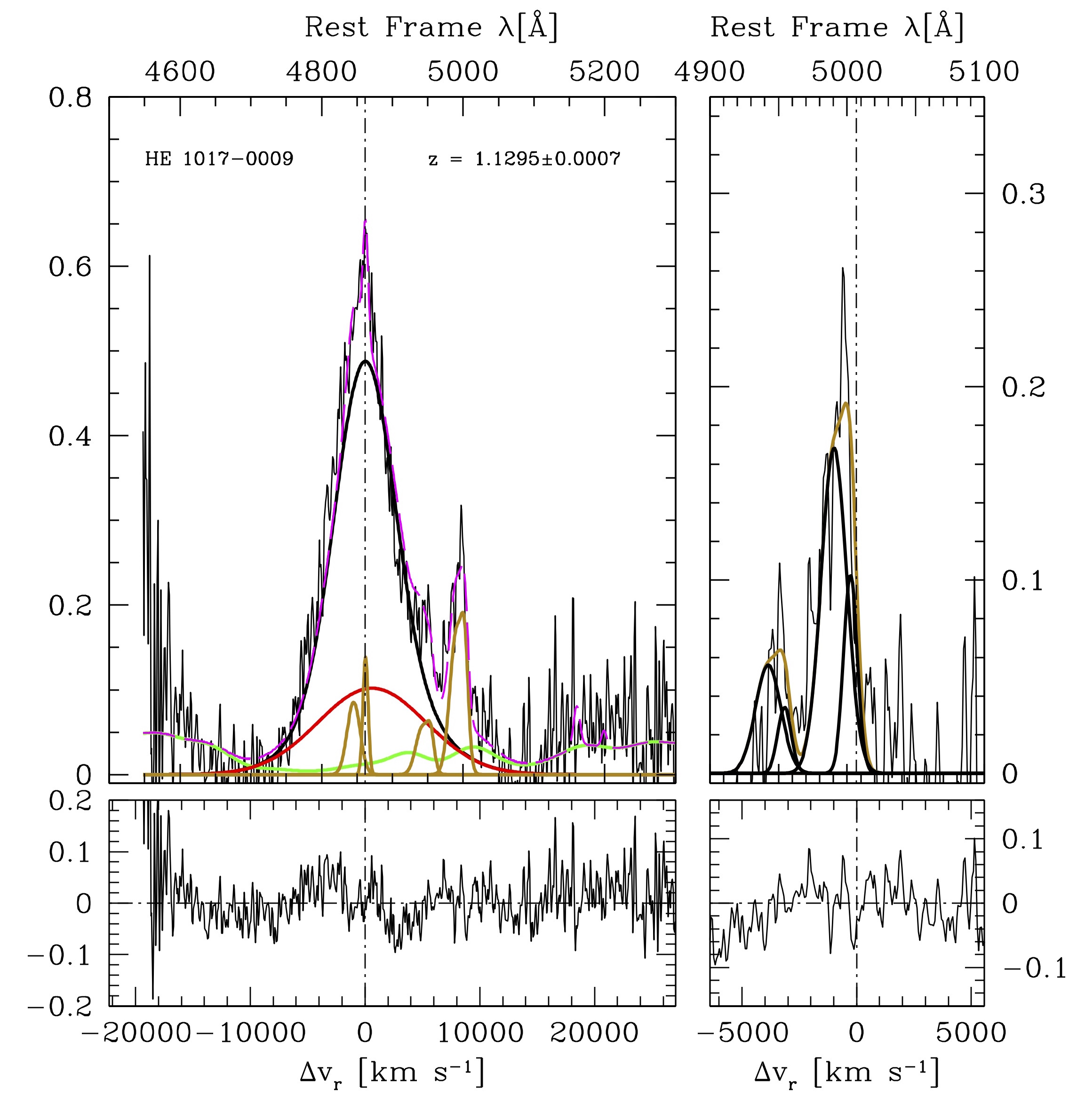}
 \vspace{0cm}
\caption{Left panel: spectrum of the Pop. B quasar HE 1017-0009 in the H$\beta$ spectral region, after continuum subtraction. Horizontal scale is rest frame wavelength in \AA\  (top) and  radial velocity shift from H$\beta$\ rest wavelength (bottom). Vertical scale is  intensity normalized to specific flux at 5100 \AA. Emission line components used in the fit are \feii\ (green), H$\beta$\ broad  (black) and very broad component (red). The \oiiiopt\ emission is traced by golden lines. Right panel: same as for left panel, after subtraction of all emission components, with only the [OIII]$\lambda\lambda$4959,5007 profiles left. 
 The two black line show the core and the blue shifted components. { Vertical dot-dashed lines trace the rest-frame wavelength of \hb\ and \oiii\ that sets the radial velocity zero points in the left and right panel respectively.} Lower panels show residual in radial velocity coordinates.  }
\label{fig:he1017}        
\end{figure}

\subsection{The case of an intermediate-$z$ \ blue outlier} 

Fig. \ref{fig:he1017} provides an example of BO in the HE sample, Pop. B source HE 1017-0009.   This source is interesting because \rfe\ and FWHM(	\hb) indicate that it is definitely Pop. B, and BOs are not observed among Pop. B sources at low $z$ and low $L$, {  while we found several cases in the HE sample. }The left panel shows the continuum subtracted H$\beta$ spectral region, and the right one provides an expansion around the [OIII]$\lambda\lambda$4959,5007 lines. In this case, the intensity of \feii\ emission is low, allowing for a reliable decomposition between the red wing of H$\beta$ and [OIII]$\lambda\lambda$4959,5007. The FWHM of the whole \oiii\ profile is 1350 \kms, with a shift at $\frac{3}{4}$\ peak intensity of $\approx$ --700 \kms. { Monte Carlo simulations of  the HE10017--0009 spectrum with S/N$\approx$20  indicate a 1$\sigma$\ error of 70 \kms\ on line peak  and $\approx 10$\% on FWHM due to noise. } The dynamical relevance of the shift i.e., the shift amplitude normalized by the line half-width half maximum \citep{marzianietal13a}, is $\Delta v(\frac{3}{4})$/HWHM $ \approx$ 1. As it is possible to appreciate from Fig. \ref{fig:he1017}, the \oiiiopt\ lines are fully blueshifted with respect to the  rest frame of the quasar. 

\subsection{Distribution of \oiiiopt\ line width and shifts}
\label{distr}
\begin{figure}[htp!]
\includegraphics[scale=0.45]{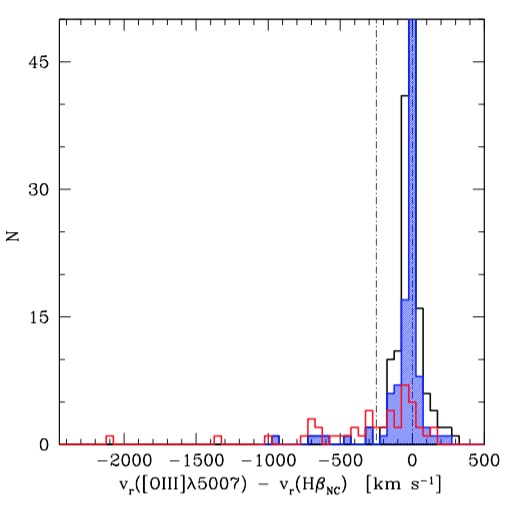} 
 \includegraphics[scale=0.45]{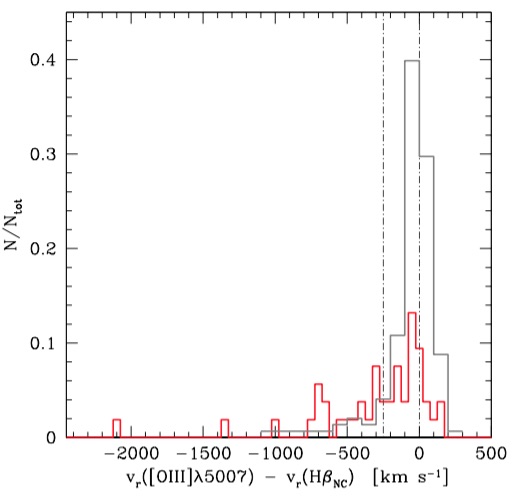}
 \caption{Distribution of \oiii\ line shifts in \kms. Top: sample of \citet{marzianietal03a} (black), with Population A sources (shaded blue) and HE  sample (red).  Bottom: HE sample (red) and \citet{bianetal05} sample (grey, bin size 100 \kms). The dashed lines at 0 \kms\ indicates the rest frame, the one at -250 \kms\ the minimum radial velocity displacement for the BOs following \citetalias{zamanovetal02}.   }
\label{fig:shifts}        
\end{figure}

At low $z$, BOs appear peculiar because of their large shifts and because of their large FWHM. Fig. \ref{fig:shifts} shows a comparison between  low $z$\ samples and the HE sample.  The top panel of Fig. \ref{fig:shifts} shows the shift distribution for the HE sample and the one of \citet{marzianietal03a} identifying Pop. A sources (blue shaded histogram). The histogram is not normalised in order to show the numbers involved in the samples. The lower panel shows the normalised distribution of \oiiiopt\ shifts for the  \citet{bianetal05} and HE samples.   
{ 
BOs are a small fraction  of the total sample at the extreme end of the shift distribution in both low-$z$\ samples.   The prevalence of BOs is about 5\%\ in \citet[][]{bianetal05}, where they might not be considered anymore statistical outliers, and somewhat  lower in \citet{marzianietal03a}, 3\%. This might not be surprising considering that  \citet{bianetal05} considered exclusively NLSy1s that, as a class, include nuclei accreting at a high rate and radiating close to the Eddington limit \citep{sulenticetal00a,mathur00,grupe04,marzianisulentic14}.  If we restrict the attention to NLSy1s the fraction of BOs in the \citet{marzianietal03a} sample increases and becomes even larger than the one of \citet[][Table \ref{tab:bo}]{bianetal05}. 

It is not surprising that there are no NLSy1s BO in the HE sample: NLSy1s,  as stressed since \citet{borosongreen92} by several authors, are not peculiar sources, but rather low-luminosity, low \mbh\ Pop. A sources, and therefore no NLSy1 is even expected in the high-luminosity HE sample. Under the assumption that all sources radiate at or below the Eddington limit, that the virial relation holds for  black hole mass estimates, and that the distance of the line emitting region can be expressed as $r_\mathrm{BLR} \propto L^{\alpha}$, then there is a minimum for FWHM as a function of $L$ that satisfies the condition \lledd = 1 and that can be written as FWHM$_\mathrm{min}(L) \propto L^{\frac{1-\alpha}{2}} $\ \citep{marzianietal09}.  As a consequence, the  Pop. A / B limit also becomes  dependent on $L$, and at very high luminosity FWHM$_\mathrm{min} \gtsim$ 2000 \kms. The $L$ dependence can be neglected for $\log L \ltsim 47$\ i.e., the condition of luminosity-independence  is roughly satisfied in  low-$z$ samples (Fig. 11 of \citealt{marzianietal09}) . 

Even in the sample of \citet{bianetal05}, or restricting to Pop. A and NLSy1s, the fraction of blue outliers remains much smaller than the one of the HE sample, $\approx 40$\%.}  At low-$z$ BOs are almost exclusively Pop. A. At high luminosity, BOs are predominantly of Pop. A but there are also several Pop. B sources (8 out of 25).  

\subsection{No ``Baldwin effect'' for the  blue outliers}

The analysis of the HE objects indicates that   \oiiiopt\ profiles like the ones shown in Fig. \ref{fig:he1017} with $\Delta v_\mathrm{r} <- 250$ \kms\ are now present in $\approx$ 40\%\ of the sample. Their equivalent width is low, typically below 20 \AA\ in the rest frame.  A fraction of sources show no detectable \oiiiopt, and these sources are typically of extreme Pop. A i.e., spectral types A3 and A4 (xA for brevity, \citealt{marzianisulentic14}). Indicatively, they are about 10 \%\ in the HE sample vs 1\% in \citet{marzianietal03a}. 

\begin{figure}[htp!]
 \includegraphics[scale=0.34]{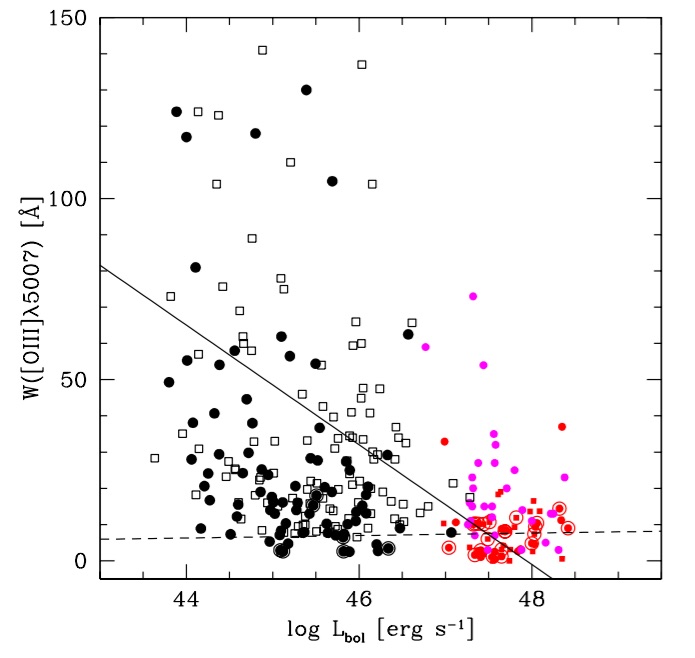} 
\caption{The \oiiiopt\ Baldwin effect. Plane W(\oiii) vs. bolometric luminosity, assumed to be 10 times the luminosity at 5100 \AA. Black symbols: sample of \citet{marzianietal03a}; red symbols: HE sample; magenta: sample of \citet{netzeretal04}. Filled circles: Pop. A  sources; open squares: Pop. B. The filled line represents an unweighted least square fit for all sources; the dashed line for BOs only (circled symbols). }
\label{fig:lew}        
\end{figure}

If the 7 blue outliers in the sample of \citet{marzianietal03a} are used as a comparison, there is no significant decrease in equivalent width with luminosity: the slope of the best fitting line is consistent with 0 within, however, a large uncertainty ($\approx 0.3 \pm 0.9$: the positive slope may even indicate an anti-Baldwin effect as found by \citealt{zhangetal13}). Considering instead the full HE and the    \citet{netzeretal04} sample at high $L$, and \citet{marzianietal03a} sample at low $L$, we see a rather strong anti correlation between \oiii\ luminosity and equivalent width (Fig. \ref{fig:lew}), with correlation coefficient $\approx$ 0.45, significant at a confidence level of 5.5 $\sigma$: it is the so-called {[\sc Oiii]} Baldwin effect \citep{brothertonetal96,baskinlaor05c}, { also found  in recent studies \citep[][]{sternlaor13,zhangetal13}}. This  {[\sc Oiii]} Baldwin effect even shows a steeper anti-correlation  than the one of the \civ\ Baldwin effect. We suggested, in several of our past works, that the \civ\ Baldwin effect, as observed in large quasar samples, could be explained as due mainly to selection effects \citep{marzianietal08} following the discovery of a strong anti correlation between W(\civ) and the Eddington ratio, stronger than the one between W(\civ) and luminosity \citep{bachevetal04,baskinlaor04}. The effect could be enhanced if the discovery of quasars is biased  toward high-Eddington  ratio sources at high $z$, as expected from flux-limited surveys.  It is impossible to define a complete sample (in terms of quasar \lledd\ and \mbh) at intermediate and high $z$ from the presently-available flux-limited samples (as shown by Fig. 2 of \citealt{sulenticetal14}). { Surveys significantly deeper than the SDSS may  affect the appearance of Fig. \ref{fig:lew}, populating the large W(\oiii) area with sources radiating predominantly at low \lledd\ that are lost in present-day surveys.} The { the Hamburg-ESO survey} can be considered flux limited, but due to the high limits in flux, it includes only part of the quasar populations. Most Pop. B sources are expected to be preferentially lost (if Pop. B sources satisfy the condition \lledd $\ltsim 0.1 - 0.2$), and this can  create a spurious L effect since Pop. B sources are the ones with the larger W(\oiiiopt). 

The BOs shift amplitudes are not strongly dependent on luminosity. If we consider the \oiii\ shifts in the samples of \citet{marzianietal03a} and HE  as a function of luminosity, we see that BOs become more frequent at high $L$, but that the data point distribution can be hardly described in terms of a correlation (Fig. \ref{fig:lsh}). The absence of a BO Baldwin effect  is also indicating  that these sources maintain similar properties at both low- and high-$z$, and that the physical mechanism behind these properties is  dependent neither on redshift nor on luminosity. We will discuss further implications in \S \ref{disc:zl}.

\begin{figure}[htp!]
 \includegraphics[scale=0.09]{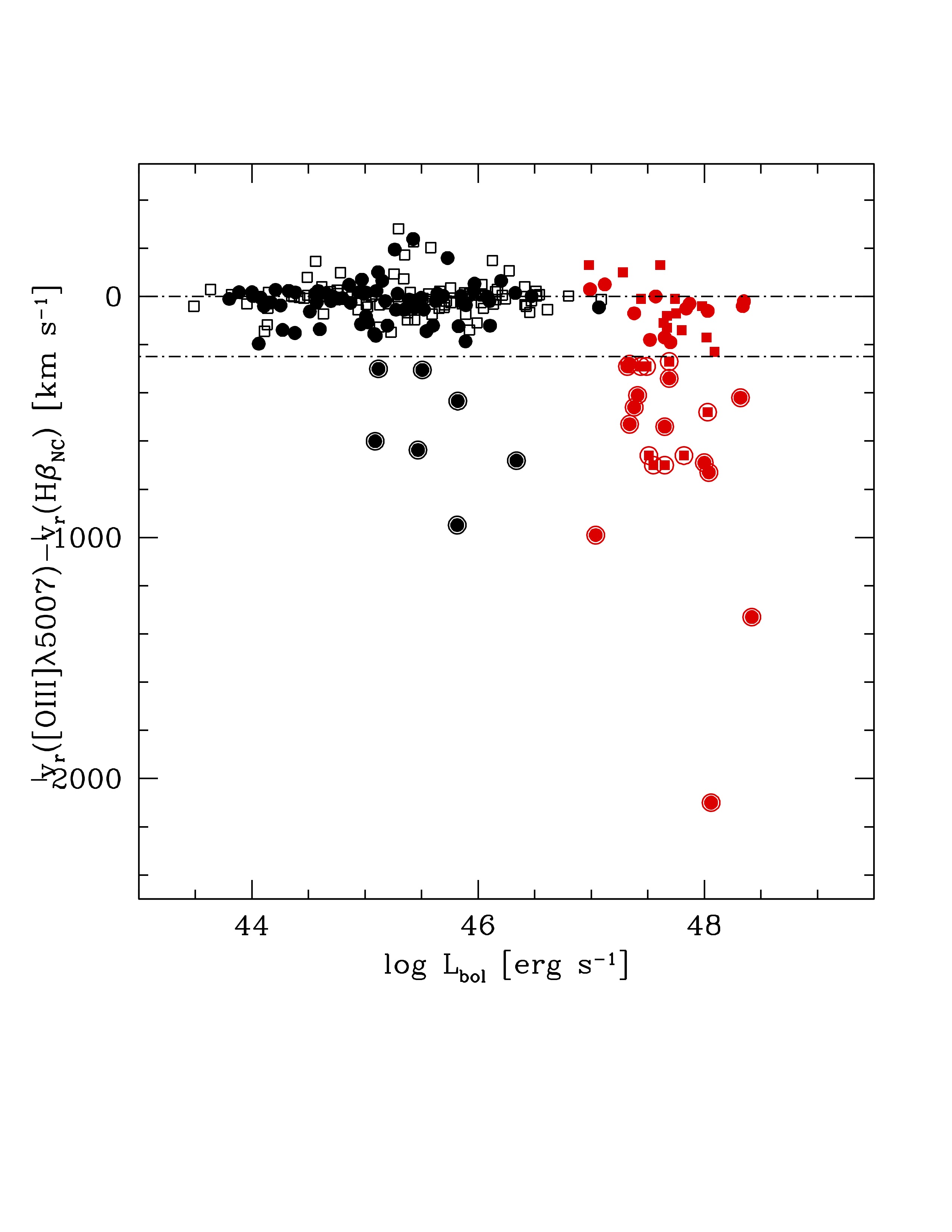} 
\caption{\oiiiopt\ line shifts vs. luminosity for the sample of \citet{marzianietal03a} and of the HE sources presented in this paper. Meaning of symbols is the same as in the previous figure. The dashed lines indicate rest frame at 0 and minimum radial velocity for BOs (-250 \kms).}
\label{fig:lsh}        
\end{figure}

\begin{figure}[htp]
 \includegraphics[scale=0.21]{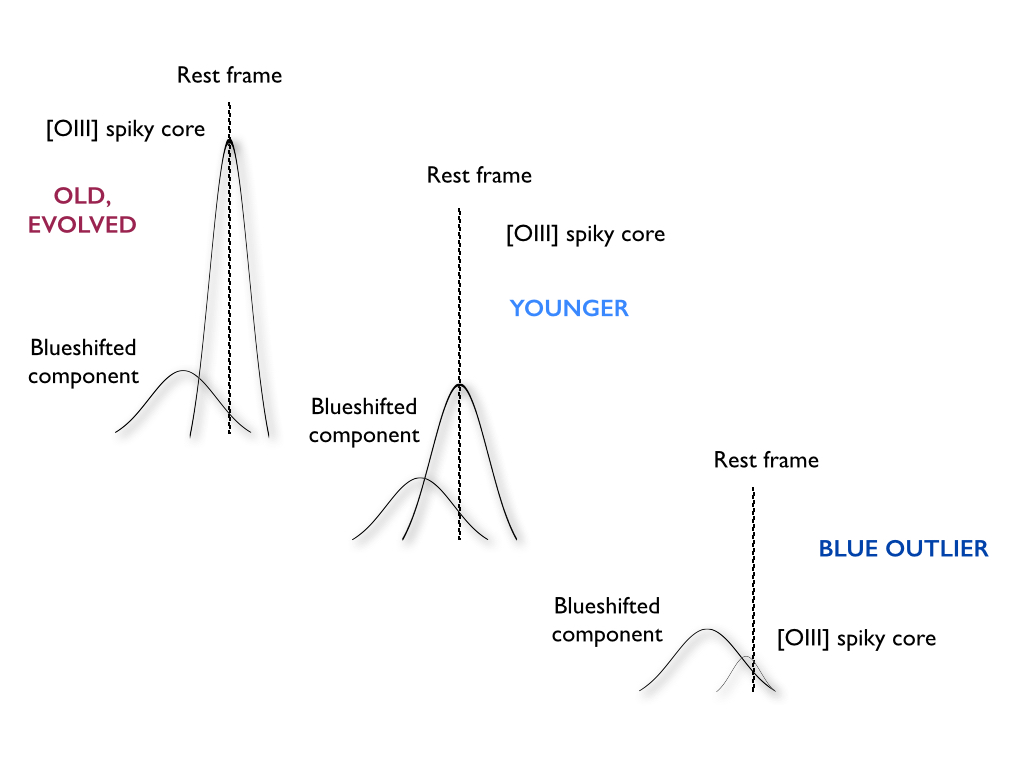} 
\caption{A sketch illustrating the change in the \oiii\ profile shape along the 4DE1 sequence. Top left: \oiiiopt\ profile with a strong, spiky core at rest frame and a blue\-shifted component that may appear as a blueward asymmetry. This profile is observed most frequently in Pop. B sources. Middle: profile more affected by the blue\-shifted component; bottom: blue outlier whose main emitter is the blueshifted component. The profile appears of low EW, semibroad and with large blueshift, and is typical  of  xA objects. { The correct interpretation of the sketch involves continuum normalized fluxes (i.e., line equivalent widths) that are approximately constant  for the blue shifted component and instead significantly decrease for the spiky core.} }
\label{fig:sketch}        
\end{figure}

\section{Discussion}
\label{disc}

\subsection{Quasar evolution or  luminosity effect?}
\label{disc:zl}

At low-$z$\ the shape of the \oiiiopt\ line profile can be well described by the relative prominence of two components:

\begin{itemize}
\item a narrower core component, typically several hundred \kms\ FWHM, spiky (at low resolution), unshifted and symmetric. The width of this line is expected to correlate with the stellar velocity dispersion of the host spheroid, $\sigma_{\star}$\ \citep{nelsonwhittle96,bonningetal05}. 
\item a broader ($\approx$ 1000 \kms\ FWHM) blueshifted component customarily referred to as the semibroad \oiiiopt\ component, and that may appear as a blueward asymmetry at low resolution, if the core component is strong. 
\end{itemize}
Fig. \ref{fig:sketch} sketches three cases with changing relative intensity of semibroad and core \oiii\ component. The top configuration is found mainly in Pop. B sources, especially if radio loud. The total W(\oiii)  is large, and the core component dominates. The middle configuration accounts for modest shifts that are found to be frequent in large quasar sample. The core component is less prominent, and a blueshifted component affects the profile. The bottom sketch provides the interpretation for the blue outliers: the blueshifted semibroad component dominates the emission. The last two configurations are found in Pop. A, with blue-outliers being increasingly more frequent toward extreme Pop. A \citep{marzianietal03c}.

At low-$z$, the shape and the trends of the \oiiiopt\ profile along the 4DE1 sequences  can be interpreted in a simple evolutionary scenario. At one extreme, we have low-mass back hole that are still accreting at a high pace. At the other end we have very massive black holes, which are, in the local Universe at low $z$, accreting at a very low rate. Some of them are massive black holes, still active (the ones that we see as extreme Pop. B) but they are probably approaching the end of their duty cycle; they have already spent most of their lifetime as luminous active nuclei.  In this sense the difference in \oiiiopt\ profile may trace an evolutionary sequence, as one goes from a compact NLR (\citealt{zamanovetal02} model suggested a size $\sim 1 $ \ pc in a local NLSy1, Ton 28) to an extended NLR, 1-10 kpc in size.  The origin of the NLR in Seyfert galaxies is associated with interaction between the radio ejecta and circumnuclear gas \citep{capettietal96}, and some evidence exists in favor of the same process for quasars \citep{leipskibennert06}. Therefore the development of a full fledged NLR may follow the same timescale of radio jet propagation on  kpc-sized scales.

There is a general consensus that the NLR gas is photoionized from the central continuum source.  The strongest evidence in favor of photoionization is provided by the constant equivalent width of the \oiiiopt\ semibroad component as a function of quasar luminosity (Fig. \ref{fig:lew}). 

At $z \approx$ 1.5, the half slit width 0.3 arcsec corresponds to a projected linear size of 2.6 kpc, assuming concordance cosmology. The inner NLR is fully included within our slit.  Even if there is probably a large scatter in the gas content of bulges in high- and low-$z$ galaxies that may account for difference in prominence of the core comment \citep{zhangetal13}, there might be a physical  limit  to the NLR size and total luminosity \citep[][]{netzeretal04,hainlineetal14}. \citet{netzeretal04} suggest that the NLR linear size cannot greatly exceed $\approx$ 6 kpc. If this is the case, at high quasar luminosity the luminosity of the semibroad blueshifted component still correlates with the quasar luminosity that is only indirectly affected by the bulge interstellar medium (ISM) distribution and physical state. In other words, at very high luminosity the semibroad blueshifted component tends to overwhelm the narrower component associated with \oiiiopt\ emission in the host spheroid. This explanation may account for the NLR ``disappearance'' revealed by \citet{netzeretal04}.  { Interestingly, if one distinguishes between sources with high and low ratio \oiii/\hb, and specifically between sources above and below $\log$ \oiii/\hbnc $= 0.5$, \citet{popovickovacevic11} found that there is a different dependence on luminosity for FWHM(\oiii) in the range $\log L_\mathrm{bol} \sim 45 - 47$ [\ergss]. If  $\log$ \oiii/\hbnc $<$ 0.5 (i.e., mostly for Pop. A sources that are in general weak \oiiiopt\ emitters), there is neat correlation between continuum luminosity at 5100 \AA\ and \oiii\ FWHM. On the converse, if $\log$ \oiii/\hbnc $>$ 0.5, the FWHM is not correlated with luminosity, and a larger fraction of sources show relatively narrow \oiii\ profiles (see Fig. 4 of \citealt{popovickovacevic11}). This finding also supports the idea that the core component -- which should dominate for $\log$ \oiii/\hbnc $>$ 0.5 and in Pop. B sources -- is not strongly related to the active nucleus. It may instead be significantly affected by host properties. } Further work should address the problem of the luminosity-size relations \citep{schmittetal03,netzeretal04,bennertetal06} in light of the two-component interpretation suggested in this and other papers.

So,  is   the increase in prevalence of blue outliers   an effect associated with luminosity or with evolution? Clearly, the HE sources are extremely luminous quasars. Unless they {\em radiate} highly super-Eddington, their black hole masses have already reached the values found for the most massive fossil black holes in the local Universe (as the one of M87, $\approx 3.5 \cdot 10^9$ \msol, \citealt{walshetal13}). They are therefore quasars whose black hole masses  are not expected to grow much more even if they are still accreting matter at high rate.  In a sense they are evolved systems, but of a kind that is not found  in the local Universe. { What at low-$z$\ can be understood as an evolutionary effect (Fig. \ref{fig:sketch}), appears as a luminosity effect at high $L$. }
The higher frequency of BOs may be  a consequence of high Eddington ratio  at high black hole mass  and of selection effects. Selection effects still  make it impossible to include low Eddington radiators even with the most massive black holes at $z \approx$ 2 in major  flux-limited surveys \citep{sulenticetal14}: they are not yet discovered because they are too faint. 

\subsection{Implication of blueshifts}

There is little doubt that the \oiiiopt\ blue\-shift is produced by Doppler effect due to gas motion with respect to the observer, and to selective obscuration, as the \oiiiopt\ emission is optically thin and scattering processes appear unlikely over the large spatial scales of \oiiiopt\ emission. From the  \oiii\ luminosity    we can retrieve information on the   {  mass of ionised gas and on the kinetic luminosity of the outflow. This is a first step in the analysis of feedback processes: mechanical feedback is possible only if there is a large flow of matter that significantly affects the ISM and star formation processes \citep[e.g.,][and references therein]{fabian12}. The following calculations are intended as order of magnitude estimates of the outflow kinetic power. } The gas mass emitting   [OIII]$\lambda$5007 can be written as, under the assumption of  constant density: 

$$
M^\mathrm{ion}_\mathrm{out} \sim 4~10^7~L_{44}({\rm [OIII]}) \left(\frac{Z}{Z_{\odot}}\right)^{-1} n_{3}^{-1}~M_{\odot}
	 \label{eq_a5}
$$

where $\frac{Z}{Z_{\odot}}$\ is the metallicity in solar units, $n$ the density in units of $10^{3}$ \cmq, and $L_{44}$\ the \oiii\ luminosity in units of 10$^{44}$ \ergss\ (two-thirds of the HE sample, and three quarters of the HE-BOs  have $\log L_{44}$(\oiii)  in the range 0.1 -- 1). The metallicity is $\frac{Z}{Z_{\odot}} \sim 0.2 - 5$\ { in the NLR of luminous quasars, and is appropriate to assume  $\frac{Z}{Z_{\odot}} \sim1$ for  density $\sim10^{3}$ \cmq, \citep{nagaoetal06b}}. The { mass outflow rate} at a distance $r$\ (1 kpc) can be written as, if the flow is confined to a solid angle of $\Omega$ of volume $V = \frac{4}{3}\pi r^{3} \frac{\Omega}{4 \pi} $:

\begin{eqnarray*}
\dot{M}^\mathrm{ion}_\mathrm{out} & = & \rho\ \Omega r^{2} v = \frac{{M}^\mathrm{ion}_\mathrm{out}}{V} 	\Omega r^{2} v \\ 
&  \approx &    135  L_{44} v_{1000} r^{-1}_{\rm 1kpc} n_{3}^{-1} \left(\frac{Z}{ Z_{\odot}}\right)^{-1} \rm M_{\odot}~yr^{-1} 
\end{eqnarray*}

where $v$\ is the line radial velocity in units of 1000 \kms. We  followed the approach of \citet{canodiazetal12}, { assuming a constant density}. Unlike \citet{canodiazetal12} however, the HE quasar outflow   is not spatially resolved and a tentative value for $r$\ has to be assumed. This value is provided by the projected linear half width  of the slit $\approx$ 2.6 kpc. The { outflow kinetic power $ \dot{\epsilon}$}  can then be estimated from: 
\begin{eqnarray*}
\dot{\epsilon} & = & \frac{1}{2} \dot{M}^\mathrm{ion}_\mathrm{out} v^{2}\\
&  \approx & 4.3 \cdot 10^{43} L_{44} v^{3}_{1000}   r^{-1}_{\rm 1kpc}  n_{3}^{-1} \left(\frac{Z}{ Z_{\odot}}\right)^{-1} \rm  erg~s^{-1}. 
\end{eqnarray*}

The  total energy  expelled over  a duty cycle { $\tau_{8}$} of 10$^{8}$ yr { (appropriate for highly accreting, massive black holes, \citealt{marconietal04})} is 

$$ \int \dot{\epsilon} dt  \sim 3.15 \cdot 10^{59 }  L_{44} v^{3}_{1000}   r^{-1}_{\rm 1kpc}  \tau_{8}  \rm~ erg.$$

This value can be compared to the  binding energy of the gas in a massive bulge/spheroid { of mass $M_\mathrm{sph}$}:

$$ U = \frac{3 G M^{2}_\mathrm{sph}f_\mathrm{g}}{5 R_\mathrm{e}} \sim 2 \cdot 10^{59} M^{2}_\mathrm{sph,11} f_\mathrm{g,0.1} R^{-1}_\mathrm{e,2.5 kpc}\rm  erg,$$

{ where $R_\mathrm{e}$\ is the effective radius in units of 2.5 kpc, and $f_\mathrm{g,0.1}$\ is the gas mass ratio in units of 0.1.}
For the  very luminous HE objects, these order-of-magnitude estimates  suggest that the outflow traced by the \oiii\ lines has a significant feedback effect on host galaxy. Unfortunately parameters like $r$ are very poorly determined; $r$ should be seen as an upper limit since the line emitting region is not resolved. This however implies that $\dot{M}^\mathrm{ion}_\mathrm{out}$\ and $\dot{\epsilon}$ are lower limits. In addition, the \oiii\ kinetic power is only part of the kinetic power, as there is gas associated with nuclear outflow that is too hot to be detected in the optical (for example, the gas associated with the ultra fast outflows (UFO) of \citealt{tombesietal10}). { A more detailed assessment of mechanical feedback effects would require knowledge of how the kinetic power of the nuclear outflow is dispersed in the host galaxy and transferred to its ISM. This is obviously beyond the scope of the present paper.   }

\section{Conclusion}

The blue outliers have been considered in a small number of works as oddities and ignored in most of AGN research. However, at low as well as at intermediate -$z$ they trace powerful outflows associated with nuclear activity that may have a significant feedback effect especially at high $L$. { An important result of the present paper is that large \oiiiopt\ blueshifts are frequent and that the \oiiiopt\ emission is more associated with the so-called semibroad component whose equivalent width is fairly constant. A straightforward implication is that the semibroad component traces an outflow that is mainly of nuclear origin. However, IFU studies of intermediate redshift quasars \citep[e.g.,][and references therein]{canodiazetal12,carnianietal15} are revealing a more complex scenario involving spatially resolved structures in which part of the \oiiiopt\ emission is associated with host galaxy star formation. All  BOs of the HE sample should be prime candidates for more detailed studies at high spatial resolution.}                                                                            


 

\newpage\eject

\vspace{1cm}

\newpage

\newpage\pagebreak

%
%

\begin{acknowledgements}
This research was supported by the Junta de Andalucõa through project TIC114,and the Spanish Ministry of Economy and Competitiveness (MINECO) through project AYA2013-42227-P.\end{acknowledgements}
\vfill\eject
\newpage\pagebreak
\bibliographystyle{spr-mp-nameyear-cnd}



\end{document}